# Study of Mass Transport in the Anode of a Proton Exchange Membrane Fuel Cell with a New Hydrogen Flow-Rate Modulation Technique

Luis Duque,[a] Antonio Molinero,[b] J. Carlos Oller,[b] J. Miguel Barcala,[b] E. Diaz-Alvarez,[a] M. Antonia Folgado,[a] and Antonio M. Chaparro*[a]

Hydrogen transport in the anode of a proton-exchange membrane fuel cell (PEMFC) has been studied with a modulation technique relating the hydrogen flow-rate ($\tilde{Q}_{H2}$) and the faradaic current ($\tilde{I}$), called *Current-modulated Hydrogen flow-rate Spectroscopy* (CH2S). A simple analytical expression for the transfer function, $H(j\omega) = n\,F\,\tilde{Q}_{H2}/\tilde{I}$, is provided, showing a skewed semicircle in Nyquist representation ($-H''$ vs. $H'$), extending from $H' = 0$ to $H' = 1$, and with the maximum frequency at $\omega_{max} = 2.33(D_{H2}/L_i^2)$, where $D_{H2}$ is the effective hydrogen diffusivity and $L_i$ the thickness of the anode gas diffusion layer (GDL). The expression for CH2S is also calculated with an existing reversible chemical reaction in the GDL. Experimental results under different operation conditions show two transport processes limiting the anode reaction, one attributed to molecular diffusion through the partially saturated GDL, and the other to the microporous layer (MPL), or its interfaces with GDL or with the catalyst layer (CL). CH2S provides the hydrogen diffusivities ($D_{H2,i}$) associated to each process under the different conditions. Current density decreases slightly the diffusivity of the GDL, while it becomes activated in the MPL; using two GDLs in the anode improves both GDL and MPL diffusivities; humidification decreases the diffusivity in both, GDL and MPL; finally, a superhydrophobic anodic CL prepared by electrospray improves hydrogen diffusivity in GDL and MPL.

## Introduction

Modulation techniques relating cell current or voltage with a variable proportional to the concentration or the flow-rate of a species, have been applied to study mass transport without interference from electrochemical kinetics.[1,2] The information is complementary to that of the electrochemical impedance spectroscopy (EIS), where mass transport effects, gathered in Warburg elements, appear overlapped in the spectra with other dynamic processes.[3,4] Among transport modulation techniques, the 'electrohydrodynamical', or 'convective-diffusion', impedance consists in modulating the rotation speed of a rotating disk and measuring the modulated current or voltage under potentiostatic or galvanostatic conditions, respectively.[2,5,6] This technique allows to determine transport properties of dissolved species (Schmidt numbers and diffusion coefficients), transport kinetics over different electrode types (uniformly accessible, partially covered, porous layers), as well as reaction mechanisms with transport limitations (electrocrystallization, corrosion). The results are normally analysed using admittance plots, i.e. magnitude and angle of the ratio response/perturbation in dependence of frequency, showing decoupled the different mass transport phenomena in the electrochemical reaction. Transport modulation techniques are also applicable in electrochemical cells with gas diffusion electrodes (GDE). In their seminal study, Grübl et al.[7] calculated the transfer function for cells having gas reservoir, diffusion layers, and active layers. The resulting Nyquist plots, i.e. $-H''$ vs $H'$, showed distinct spiral-shaped features, more sensitive to mass transport conditions than conventional EIS.

Of high practical interest is the application of these modulation techniques to proton-exchange membrane fuel cells (PEMFCs), where mass transport in gas diffusion electrodes (GDEs) and flow-field plates become rate determining steps at intermediate and high current densities, so they pose an important limitation to attain high power densities. Sorrentino et al.[8,9] combined the electrical response of a PEMFC with a periodic perturbation of reactants concentrations (*concentration-alternating frequency response*) to study dynamics of reactants and water in the cathodic layers and in the channels of the flow-field plate. Pressure modulation in PEMFC cathodes was used by Shirsath et al.[10,11] to analyze the effect of humidification on oxygen transport in the cathode, and by Zhang et al.[12] to study the transport resistance of the flow-field.

[a] L. Duque, Dr. E. Diaz-Alvarez, Dr. M. A. Folgado, Dr. A. M. Chaparro
Low Temperature Fuel Cell Group, Department of Energy
CIEMAT (Centro de Investigaciones Energéticas, Medioambientales y Tecnológicas)
Avda. Complutense 40, Madrid 28040, Spain
E-mail: antonio.mchaparro@ciemat.es

[b] A. Molinero, Dr. J. C. Oller, Dr. J. M. Barcala
Microelectronic Unit, Department of Technology
CIEMAT (Centro de Investigaciones Energéticas, Medioambientales y Tecnológicas)
Avda. Complutense 40, Madrid 28040, Spain

Water mass transport has been studied with neutron images in a PEMFC under alternate current perturbation to obtain information on local water exchange and generation.[13]

So far, most of the reported studies using transport modulation techniques in PEMFC are dedicated to the study of oxygen transport in the cathode, while the anode is considered a silent counter-electrode. Such an assumption is reasonable under normal operation conditions, where oxygen kinetics and transport losses limit the cell response with respect to the fast hydrogen kinetics. However, testing hydrogen transport in the anode of a PEMFC is useful for different reasons. Due to its fast electrochemical kinetics, hydrogen oxidation is mostly transport limited in the porous media of a gas diffusion electrode,[14,15] which makes a PEMFC a good platform for fundamental transport studies. In addition, the possibility to monitor in-operando anode transport properties in a PEMFC may help to identify phenomena like anode flooding and hydrogen starvation that lead to cell degradation.[16–18] Anode transport analysis also helps to optimize new electrode and flow fields, and porous layers properties.[19–21]

For the study of anode transport properties in a PEMFC, we have recently developed a flow-modulation technique relating the hydrogen flow-rate in the anode with the current, called *Current-modulated Hydrogen flow-rate Spectroscopy* (CH2S).[22,23] In CH2S, the perturbation is the cell current and the response is the flow-rate of hydrogen, with the transfer function (*H*) given by:[23,24]

$$H(j\omega) = nF\frac{\widetilde{Q_{H2}}}{\tilde{I}}, \quad (1)$$

where $n$ ($=2$) is the electronic exchange number, $F$ ($= 96485$ C mol$^{-1}$) the Faraday constant, $\tilde{Q}_{H2}$ the perturbed hydrogen flow-rate, and $\tilde{I}$ the applied current perturbation. When CH2S is measured under closed anode (dead-end mode), $\tilde{Q}_{H2}$ is identical to the (alternate) faradaic flow rate, hence $|H(j\omega)|$ evolves from 0 to 1 when sweeping from high to low frequencies under normal conditions. At intermediate frequencies, CH2S shows the relaxation hydrogen transport phenomena in the porous layers of the anode, i.e. gas-diffusion layer (GDL), microporous layer (MPL), and catalyst layer (CL), whose time constants are higher than the experimental time response.[23] Such phenomena comprise principally the molecular diffusion through the partially water-saturated porous layers.[14,25–27] In the CL, transport kinetics is more complex, since molecular diffusion coexists with Knudsen diffusion in smaller pores, and with transport through the ionomer film covering the catalyst particles.[28,29] The latter gives rise to the so called '*local transport resistance*', dependent on porosity, ionomer contents, and catalyst load, and includes processes like the dissolution and transport of gas molecules in the ionomer, and their transport and possibly surface diffusion towards catalyst centers.[30–34] The transport resistances have been studied experimentally with EIS,[35,36] hydrogen limiting current,[37,38] and CO masking,[39,40] however the nature of the transport processes, especially those in the CL, is not well understood. CH2S may contribute to the better understanding of transport processes of gases in porous electrodes of PEMFCs.

In this work, the CH2S technique is used to study the mass transport properties of PEMFC anodes. An analytical expression for the transfer function in Eq. 1 is calculated for the anodic GDL (Figure 1a), including an expression with hydrogen reversible kinetics in the GDL (Figure 1b). In the second part, experimental CH2S results are shown varying basic experimental conditions in a PEMFC, like current density, two GDLs, humidification, and CL hydrophobicity. The results provide the effective diffusivity of the GDL under different experimental conditions, and reflect some transport properties of the other anode layers.

## Theoretical CH2S Transfer Function

The expression for the transfer function of a GDL is calculated assuming a one-dimensional, continuous porous layer, as shown in Figure 1a, where hydrogen transport is characterized by an effective diffusion coefficient ($D_{H2}$) according to the second Ficks' law:

$$\frac{\partial c_{H2}}{\partial t} = D_{H2}\left(\frac{\partial^2 c_{H2}}{\partial x^2}\right), \quad (2)$$

where $c_{H2}$ is the hydrogen concentration. The one-dimensional treatment neglects variations of $c_{H2}$ in the PEMFC cell plane. Under a sinusoidal perturbation the transport equation can be applied to the $c_{H2}$ phasor, i.e. $\widetilde{c_{H2}}$, so Eq. 2 can be written as:

$$j\omega\widetilde{c_{H2}} = D_{H2}\frac{\partial^2 \widetilde{c_{H2}}}{\partial x^2}, \quad (3)$$

where $j$ is the imaginary unit and $\omega$ the angular frequency. Eq. 3 is solved analytically assuming finite-length linear diffusion in the GDL,[3] limited by the CL on one side and by a static

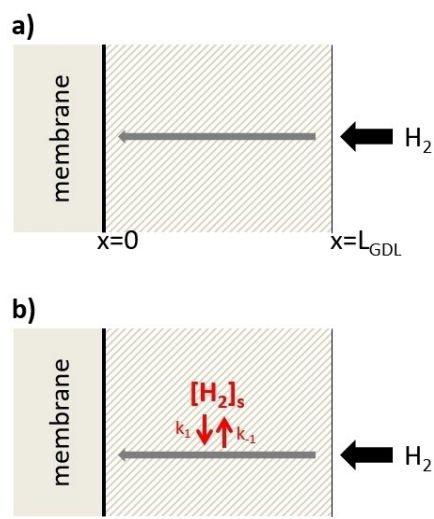

**Figure 1.** Schemes of hydrogen transport in the electrode, without (a), and with (b) reversible chemical kinetics in the GDL.

hydrogen atmosphere on the other side, with the GDL thickness $L_{GDL}$, thus the boundary conditions remain:

$$x = 0 : D_{H2}\left(\frac{\partial \widetilde{c_{H2}}}{\partial x}\right) = -\frac{\widetilde{I}}{nFA}, \quad (4)$$

$$x = L_{GDL} : \widetilde{c_{H2}} = 0, \quad (5)$$

Semi-infinite boundary conditions could also apply, depending on the thickness of the layer and diffusivity, leading to a slightly different solution (see Supporting Information).

The boundary conditions in Eqs. 4 and 5 pose some experimental restrictions. Hydrogen flow-rate ending in faradaic current generation at the CL (x=0) (Eq. 4) requires a dead-end anode, while no concentration modulation outside the GDL (Eq. 5) requires the back surface of the anode to be in contact with a sufficiently large, static, hydrogen supply. On the other hand, electrochemical and transport resistances in the CL are not considered, which is acceptable for thin layers and fast hydrogen kinetics.

The resulting perturbed concentration profile in the GDL is:

$$\widetilde{c_{H2}}(x) = \frac{\widetilde{I}}{nFA\sqrt{j\omega D_{H2}}} \frac{e^{-x\sqrt{\frac{j\omega}{D_{H2}}}}\left(e^{2L_{GDL}\sqrt{\frac{j\omega}{D_{H2}}}} - e^{2x\sqrt{\frac{j\omega}{D_{H2}}}}\right)}{1 + e^{2L_{GDL}\sqrt{\frac{j\omega}{D_{H2}}}}}, \quad (6)$$

and the perturbed flow rate at $x = L_{GDL}$:

$$\widetilde{Q_{H2}} = D_{H2}\left(\frac{\partial \widetilde{c_{H2}}}{\partial x}\right)_{x=L_{GDL}} = \frac{\widetilde{I}}{nFA}\,\mathrm{sech}\left(L_{GDL}\sqrt{\frac{j\omega}{D_{H2}}}\right), \quad (7)$$

from which the expression of the transfer function (Eq. 1) results:

$$H(j\omega) = nFA\frac{\widetilde{Q_{H2}}}{\widetilde{I}} = \mathrm{sech}\left(L_{GDL}\sqrt{\frac{j\omega}{D_{H2}}}\right). \quad (8)$$

Nyquist and Bode representations of Eq. 8 are shown in Figure 2. The Bode representation (Figure 2a) shows the module and angle evolution with frequency, reflecting modulation absorption by hydrogen transport in the GDL. The Nyquist plot (Figure 2b) shows a skewed semicircle, starting at the origin at high frequency, and ending at low frequency at the real unit value ($H' = 1$, $H'' = 0$) which corresponds to 100% faradaic flow-rate modulation. The frequency at the maximum is related to the diffusivity and thickness of the GDL according to:

$$\omega_{max} = 2.33\frac{D_{H2}}{L_{GDL}^2}. \quad (9)$$

Similar result was obtained by Grübl et al.[41], using numerical simulation, but with a proportionality factor of 2.93, whereas the EIS finite-length Warburg impedance with transmissive boundary results in the factor 2.54,[4]. For the semi-infinite solution, the proportionality factor is 1.23 ($= \pi^2/8$, see Supp. Inf.).

*Reversible chemical kinetics in the GDL.* The possibility of a reversible reaction of hydrogen inside the GDL before oxidation at the CL, as depicted in Figure 1b, refers to adsorption/desorption processes on pores walls, or the formation of a reversible compound. In this case, the transport equation is:

$$\frac{\partial c_{H2}}{\partial t} = D_{H2}\left(\frac{\partial^2 c_{H2}}{\partial x^2}\right) - k_{-1}c_{H2} + k_1 c_{H2,s}, \quad (10)$$

where $k_1$ (s$^{-1}$) and $k_{-1}$ (s$^{-1}$) are hydrogen release and capture reaction constants, respectively, and $c_{H2,s}$ (mol cm$^{-3}$) is the captured concentration. The case is similar to the Gerischer finite-length impedance with homogeneous reaction,[3] and can be solved in a similar way by using the boundary conditions in Eqs. 4 and 5 (see Supporting Information, solved with semi-infinite and finite boundaries), resulting in (finite boundaries):

$$\widetilde{c_{H2}}(x) = \frac{\widetilde{I}}{nFA\sqrt{j\omega D_{H2}}} \frac{1}{k_{-1} + k_1}$$
$$\left(k_1 \frac{e^{-x\sqrt{\frac{j\omega}{D_{H2}}}}\left(e^{2L_{GDL}\sqrt{\frac{j\omega}{D_{H2}}}} - e^{2x\sqrt{\frac{j\omega}{D_{H2}}}}\right)}{1 + e^{2L_{GDL}\sqrt{\frac{j\omega}{D_{H2}}}}} + \right. $$
$$\left. k_{-1}\frac{e^{-x\sqrt{\frac{j\omega}{D_{GDL}} + \frac{k}{D_{GDL}}}} - e^{x\sqrt{\frac{j\omega}{D_{GDL}} + \frac{k}{D_{GDL}}}}}{1 + e^{-2GDL\sqrt{\frac{j\omega}{D_{GDL}} + \frac{k}{D_{GDL}}}}}\right) \quad (11)$$

$$H(j\omega) = nFA\frac{\widetilde{Q_{H2}}}{\widetilde{I}} = \frac{1}{k_{-1} + k_1}$$
$$\left(k_1 \mathrm{sech}\left(L_{GDL}\sqrt{\frac{j\omega}{D_{H2}}}\right) + k_{-1}\mathrm{sech}\left(L_{GDL}\sqrt{\frac{j\omega + k_{-1} + k_1}{D_{H2}}}\right)\right). \quad (12)$$

Eq. 12 for the transfer function with chemical kinetics shows similar representation and frequency at maximum as in the absence (Eq. 9), as shown in Figure 2. The most significant change is in the low frequency limiting value, which is now dependent on the kinetics constants $k_1$ and $k_{-1}$. Two limiting situations can be considered:

$$k_{-1} + k_1 \ll D_{H2}/L_{GDL}^2 \rightarrow H'(\omega = 0) = 1, \quad (13)$$

$$k_{-1} + k_1 \gg D_{H2}/L_{GDL}^2 \rightarrow H'(\omega = 0) = k_1/(k_{-1} + k_1). \quad (14)$$

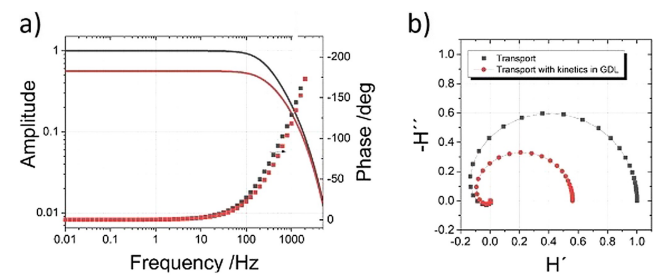

**Figure 2.** a) Bode and b) Nyquist representations of CH2S for the two cases in Figure 1, ie. in the absence (black) and presence (red) of reversible chemical kinetics in the GDL.

Hence, a reversible chemical kinetics in the GDL will cause a decrease in the low frequency real value (below one) of the CH2S, which will be significant with thick-enough GDLs, low hydrogen diffusivity, and when hydrogen capture is faster than its release ($k_{-1} \gg k_1$).

## Experimental

A conventional single PEMFC of 15.2 cm$^2$ active area was used for CH2S experiments, with gold-coated stainless-steel flow-field plates of 2 mm thickness having double-serpentine channeling (1 mm×1 mm channel section, and 1 mm rib). The end-plates were stainless-steel plates of 1.4 cm thickness with integrated heating resistors for thermal control, and clamped with eight screws tightened to a controlled torque (2 Nm). Commercial GDEs (FCSTORE, 0.25 mg Pt·cm$^{-2}$; 30 wt% ionomer; 10 μm CL thickness; 300 μm woven carbon cloth GDL) were used as anodes and cathodes, and Nafion 212NR polymer (Ion Power) as membrane. A cross section image of the MEA is shown in Supplementary Information, Figure S6. Polarization curves were obtained with a test bench under control of temperature, gas flow rate, humidification, and back-pressure. Oxygen (Air Liquide, 99.995%, 1 bar$_g$, $\lambda$=3.0) and dry hydrogen (Air-Liquide 99.999%, 1 bar$_g$) were used in cathode and anode, respectively.

For CH2S measurements, the same test bench was used, with additional elements as shown in the scheme of Figure 3. Photographs of the components are in Figure S2 of Supporting Information. The measurements were taken under dead-end anode (i.e. $\lambda$=1), which is a premise of the theory exposed in the previous section. The hydrogen flow-rate, entirely due to generation of faradaic current, is measured with a microbridge sensor (Honeywell, AWM3100V) at the inlet port of the anode. The working principle of this sensor is based on the measurement of the rate of heat transfer from a heater resistor to a temperature-sensing resistor located on both sides of the heater. The sensor has a time response of 3 ms according to the manufacturer. The linearity and time response of the sensor were checked with a PEMFC operated in dead-end mode[23] (See Supp. Inf., Figures S3 and S4). The fuel cell in dead-end mode was used for the calibration of the flow-meter according to the Faraday equation. Calibration results were checked once with a glass soap-bubble flow-meter. A sinusoidal current modulation with 50 mA peak-to-peak (3 mA cm$^{-2}$) was superposed to the dc current load by means of the ac-generator of a potentiostat/galvanostat (Autolab 30 N, Metrohm) connected to the external control of an electronic load (Kikusui PLZ164WA). Spectra were taken from 1 kHz to 0.01 Hz, and the transfer function ($H$) was obtained with the frequency response analyzer of the potentiostat by connecting the flow-meter signal to the X-port, and the load current signal to the Y-port.

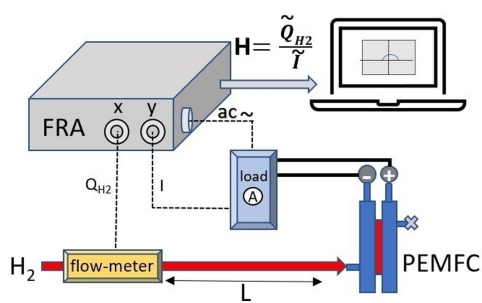

**Figure 3.** Scheme of the experimental set-up for CH2S.

## Results and Discussion

The magnitude and angle of the CH2S spectra at four cell current densities are shown in Figure 4 (see polarization curve in Figure S5 of Supporting Information). The magnitude evolves from zero to one when going from high to low frequency, as expected, except for the lowest current where the magnitude attains a lower value, 0.9. At least three processes with relaxation times ($\tau_i = 1/\nu_i$) can be identified, as indicated in the figure. In order to measure their relaxation times, the experimental curves are fitted to a weighted addition of theoretical curves, according to:

$$H(j\omega) = \sum_i w_i H_i(j\omega), \tag{15}$$

where $H_i(j\omega)$ is the transfer function (Eq. 8) of the process $i$, and $w_i$ its relative weight ($\sum_i w_i = 1$). Such analysis assumes identical boundary conditions for each layer, which is an approximation. More rigorous treatments of multiple serial transport layers can be found in the literature.[42,43] The fitted curves are plotted in the same figure, and resulting parameters are given in the bar-graphs of Figures. 4b—e. The high frequency relaxation ($\tau_1$) is in good accordance with the flow-meter time response of 4 ms (Supporting information, Fig-ure S4), and remains constant at any operation condition, thus it can be attributed to the response rate of the experimental set-up.[44] At lower frequency, a second relaxation time, $\tau_2$, is assigned to the anodic GDL according to previous works,[44]. The evolutions of $\tau_2$ and $w_2$ with the current density are shown in Figs. 4b and c. Increasing the current density causes a slight increase in $\tau_2$, which reflects some decrease in hydrogen diffusivity ($D_{H2,GDL} = L_{GDL}^2/\tau_2$). Using the anodic GDL thickness $L_{GDL} = 250$ μm, values of the hydrogen effective diffusivities are given in Table S1 of Supporting Information. They vary in the interval 0.006 cm$^2$ s$^{-1}$ < $D_{H2,GDL}$ < 0.012 cm$^2$ s$^{-1}$, well in the range for this type of GDL,[45] with mass transport resistance ($R_{GDL} = L_{GDL}/D_{H2,GDL}$) in the interval 3 s cm$^{-1}$ > $R_{GDL}$ > 1.5 s cm$^{-1}$, in good agreement with results obtained with the limiting current technique.[45,46]

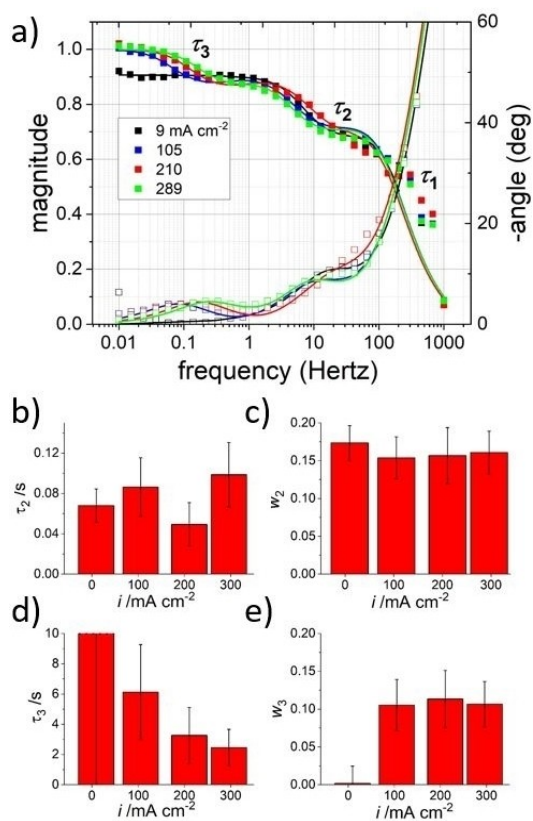

**Figure 4.** a) CH2S magnitude and angle at four current densities. Cell conditions: feeding with hydrogen in anode (dead-end), and oxygen ($\lambda$=3.0). 50 °C, 0.8 bar, 0% RH. Solid lines are fitted curves according to Eq. 15. b)–d) Evolution of $\tau_2$ (b) $w_2$ (c), $\tau_3$ (d) and $w_3$ (e) with the current density, from the analysis of results in Figure 4a.

The effect of the current density on the effective hydrogen diffusivity in the GDL can be attributed to changes in water saturation (s). The slight increase observed in Fig. 4 can be explained from the dependence of the diffusivity on s, according to an expression of the type:[28]

$$D_{H2,GDL} = D_{H2,0}^{3.5} \varepsilon^{3.5}(1-s)^{2.15}, \quad (16)$$

where $D_{H2,0}$ (= 0.75 cm$^2$ s$^{-1}$ for hydrogen at 1 atm and 295 K[47]) is the open space diffusivity of hydrogen, and $\varepsilon$ (=0.70) is the average, compressed, GDL porosity. Resulting saturation values fall between $s=0.21$ at low current density to $s=0.77$ at higher current, higher than normally found, $s<0.5$, that could be due to more water accumulation by the dead-end anode operation conditions used here. The determined water saturations are average values, taking into account the water profile that develops in the GDL in in-plane and through-plane directions in operando cells.[48–51]

The low frequency relaxation signal in Figure 4, with time constant $\tau_3$, shows a clear trend with the current density, as shown in Figures 4d and e, reflecting a hydrogen transport process of different characteristics than $\tau_2$. Hydrogen transport rate increasing with the current density cannot be explained by the increasing water saturation, as was inferred for the anodic GDL. Such opposed evolution may reflect a thermal activation of hydrogen transport caused by the ohmic resistance to current flow. Such thermal effect could also activate a hydrogen release process ($k_1$ in Figure 1b) responsible for the module below unit at the lowest current density (cf. Figure 2). More information on this process requires more experimentation changing basic parameters of the transport layers.

One principal parameter is the GDL thickness, which has a direct impact on the hydrogen transport rate measured with CH2S response. Figure 5a compares CH2S response of cells mounted with one and two GDLs in the anode, and the results of the analysis are given in Figures 5b-e, at two current densities.

The time constant ascribed to the GDL ($\tau_2$) increases with two GDLs (Figure 5b), and in similar extent at the two current densities, as expected from having a thicker GDL. However, the calculated effective diffusivity of the double GDL is higher (see Table S1 of Supporting Information), which must be a result of a lower saturation than with a single GDL. Doubling the anode GDL has been modelled by Medici et al. using a coupled continuum-pore network model.[52] They found that double GDL in anode gives rise to a decrease in water contents and enhances water transport and heat dissipation towards the cathode. Such result is in good accordance with the increase in effective diffusivity of the double GDL observed here. The third time constant ($\tau_3$) decreases by adding the second anodic GDL (Figure 5d), which confirms the different nature of this hydrogen transport process anticipated from the previous results in Figure 4, beeing effectively accelerated by adding a second GDL. Modelling results[52] explain such effect as a result of the thermal activation of hydrogen transport in the MPL, which is a plausible hypothesis for $\tau_3$.

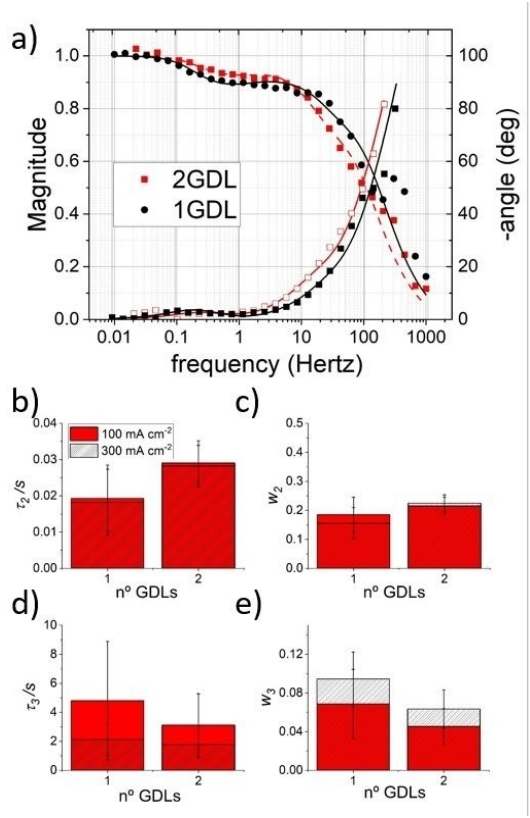

**Figure 5.** a) Bode plots of CH2S with one and two anodic GDLs, with fitted curves according to Eq. 15. Cell conditions: feeding with hydrogen in anode (dead-end), and oxygen ($\lambda=3.0$, 100% RH). 30 °C, 0.8 bar, 300 mA cm$^{-2}$. b)–e) Evolution of $\tau_2$ (b), $w_2$ (c), $\tau_3$ (d), and $w_3$ (e), for cells with one and two GDLs in the anode, and at two current density. From analysis of results in Figure 5a.

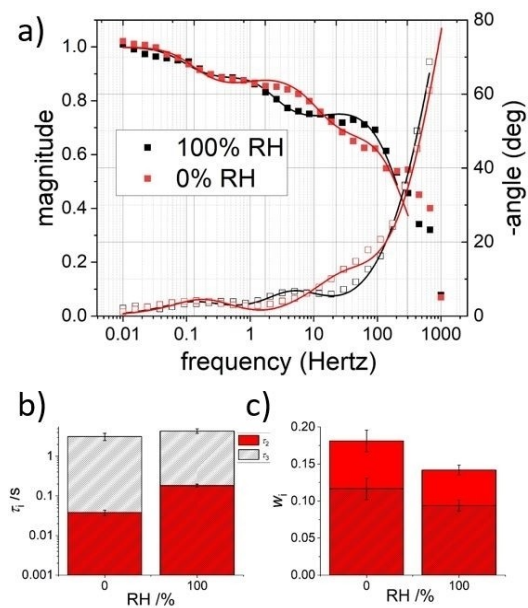

**Figure 6.** a) Comparing CH2S with 0% RH and 100% RH. Cell temperature 50 °C, current density 200 mA·cm$^{-2}$. b) Time constants (notice logarithmic scale), and c) weight factors.

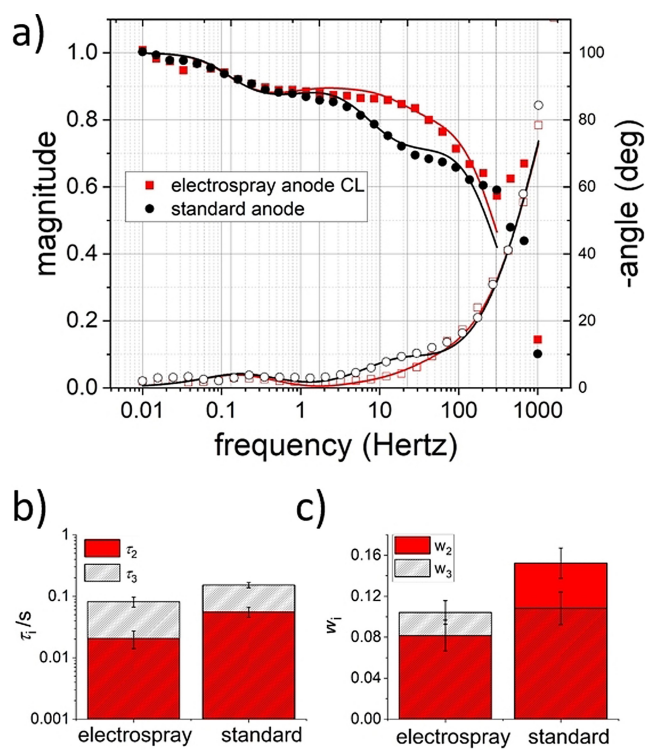

**Figure 7.** a) CH2S of cells with anodic electrosprayed (superhydrophobic) CL, and conventional CL. Cell temperature 50 °C, current density 200 mA·cm$^{-2}$. b) Time constants of the two CH2S signals (logarithmic scale), and c) weight factors.

The effect of hydrogen humidification on the CH2S signal is shown in Figure 6a. Measurements at 100 % and 0 % RH were obtained after operating the cell during two hours under 100 % humidification and without humidification, respectively, with other cell conditions 50 °C, $\lambda_a/\lambda_c = 1.5/3.0$, and 200 mA cm$^{-2}$. The time constants $\tau_2$ and $\tau_3$ increase with humidification (Figure 6b) which reflects the larger water saturation in the pores. Also the weight factors decrease with humidification for both processes. At the lowest frequency signal, $\tau_3$, humidification causes some more complex change, with the probable presence of two signals that are more evident in the Nyquist representation (see Supporting information, Figure S7). It is probable that high humidification is causing liquid water accumulation in different parts of the anode porous layers.

Mass transport properties of a PEMFC electrode change by using a superhydrophobic CL due to increasing the water transport rate.[53,54] CH2S results in Figure 7 compare a cell with a commercial anode with a cell having electrosprayed CL with superhydrophobic character. Important differences in the CH2S response are observed at the frequency range of the GDL relaxation ($\tau_2$), whereas some minor changes are observed at lower frequency ($\tau_3$). The analysis (Figure 7b) shows lower $\tau_2$, i.e. faster hydrogen transport, in the GDL in contact with superhydrophobic CL, in agreement with the decrease observed in the mass transport resistance.[54]

In summary, CH2S technique is able to show the characteristics of hydrogen transport in the anode of a PEMFC under different operation conditions. It allows the identification of two rate limiting processes in the frequency range available. The one at intermediate frequency can be assigned to molecular transport in the GDL. Taking $L_{GDL} = 250$ μm (Figure S6), the effective diffusivities ($D_{GDL} = L_{GDL}^2/\tau_2$) under the conditions used in this work are in the range 0.3 10$^{-2}$ cm$^2$ s$^{-1}$–3 10$^{-2}$ cm$^2$ s$^{-1}$ (Table S1 of Supporting Information). The low frequency transport process has different nature. Its CH2S signal shows hindering by water contents (effects of humidification in Figure 6), but acceleration by GDL thickness and current density. At the same time, an electrosprayed CL shows small effect on the parameters of this signal (Figure 7), which appears more related with the MPL, and/or its interfaces with the other layers MPL-GDL and MPL-CL. Taking $L_{MPL} = 70$ μm (Figure S6), the calculated effective diffusivities ($D_{MPL} = L_{MPL}^2/\tau_3$) under the conditions used here are in the range 0.8 10$^{-5}$ cm$^2$ s$^{-1}$– 6.3 10$^{-5}$ cm$^2$ s$^{-1}$ (Table S1 of Supporting Information). More experimentation is in progress to support the assignation of this signal.

## Conclusions

*Current Modulated Hydrogen flow-rate Spectroscopy* (CH2S), a transport impedance technique, has been applied to the study of transport properties of the anode of a PEMFC. An analytical model is solved to obtain the expression of the dependence of the transfer function on porous layer properties, i.e. the thickness and the effective hydrogen diffusivity. The possibility of a reversible chemical reaction in the porous layer is also accounted for.

CH2S results are shown under different experimental conditions (current density and humidification), and anode characteristics (GDL thickness and superhydrophobic CL). Two dynamic processes are identified related to hydrogen transport in the anode. One, with time constant ($\tau_2$) from 0.01 s to 0.05 s, is attributed to transport in the partially-saturated GDL, resulting in effective diffusivities from 0.3 10$^{-2}$ cm$^2$ s$^{-1}$ to 3 10$^{-2}$ cm$^2$ s$^{-1}$. And another one, with time constant $\tau_3 > 1$ s, whose origin is probably related to the MPL, or interfacial layers (MPL-GDL or MPL-CL), resulting in diffusivities from 0.8 10$^{-5}$ cm$^2$ s$^{-1}$ to 6.3 $10^{-5}$ cm$^2$ s$^{-1}$. The CH2S technique can be used to measure in-operando effective diffusivities of anodes in PEMFCs.

## Acknowledgements

Funding by PORHYDRO2 project (TED2021-131620B-C22, 'Preparation and characterisation of catalyst layers fabricated by electrospray for proton exchange fuel cells with passive gas feed'), and ELHYPORT project (PID2019-110896RB-I00, 'Hydrogen fuel cells with advanced membrane-electrode assemblies for their integration in low power and portable applications'), by the Spanish Ministry of Science and Innovation.

## Conflict of Interests

The authors declare no conflict of interest.

## Data Availability Statement

The data that support the findings of this study are available from the corresponding author upon reasonable request.